\documentclass[letterpaper, 12pt]{article}
\def\MYTITLE{What's the Magic Formula Instrument?}
\def\MYKEYWORDS{instrumental variables, formula instruments, sensitivity analysis}
\usepackage{atpaper}
\usepackage{atmath}
\usepackage{caption}
\newcommand{\betariv}{\hat{\beta}_{\textsc{riv}}}
\newcommand{\betalb}{\underline{\beta}_{\textsc{riv}}}
\newcommand{\betaub}{\overline{\beta}_{\textsc{riv}}}

\newcommand{\ma}{\textsc{ma}}
\newcommand{\pop}{\textsc{pop}}

\newcommand{\PJ}{\mathcal{P}_{\textsc{j}}}
\newcommand{\PM}{\mathcal{P}_{\textsc{m}}}

\newcommand{\pbh}{\bar{p}_{\textsc{bh}}}
\newcommand{\qbh}{\bar{q}_{\textsc{bh}}}

\title{\MYTITLE%
  \thanks{
    We thank Steven Durlauf, Magne Mogstad, Guillaume Pouliot, Zeyang Yu, and participants in the Harris PhD workshop for helpful comments.
  }
}
\author{
  Peizan Sheng\thanks{Harris School of Public Policy, University of Chicago.}
  \and
  Alexander Torgovitsky\thanks{Kenneth C. Griffin Department of Economics, University of Chicago.}
}

\begin{document}

\maketitle
\thispagestyle{empty}

\abstract{
  Two recent papers by \cite{borusyakhull2023e, borusyakhull2026e} propose using known formulas to adjust linear instrumental variable estimators for confounding covariates.
  Implementing this ``formula instrument'' approach requires making a parametric assumption on the distribution of the unobserved shocks that generated the instrument.
  We develop a method for systematically evaluating the sensitivity of formula instrument estimates to this parametric assumption.
  The method is straightforward to implement using our companion \texttt{R} package \texttt{formulaiv}.
  We use our method to reanalyze the applications in both \cite{borusyakhull2023e} and \cite{borusyakhull2026e}.
  In both applications, we find that a variety of estimates of different signs and magnitudes can be recovered by slightly changing the shock distribution.
}

\vspace*{\fill}
\newpage

\onehalfspacing

\section{Introduction}
\label{sec-intro}

In two recent papers, \cite{borusyakhull2023e, borusyakhull2026e} consider causal inference strategies based on ``formula instruments,'' where an instrumental variable (IV) is created by applying a known formula to other variables.
Similar approaches have long been used for policy eligibility \citep{curriegruber1996tqjoe}, tax liability \citep{grubersaez2002jope}, and the ubiquitous \cite{bartik1991} (or shift-share) instruments \citep{blanchardkatz1992bpoea, goldsmith-pinkhamsorkinswift2020aera}.
\cite{borusyakhull2023e} take this idea a step further by proposing that researchers specify the entire data generating process for the formula instrument, including the ex-ante distribution of counterfactual shocks that produces its exogenous variation.

In this paper, we develop a method for conducting sensitivity analysis to this assumed distribution of shocks.
As \citet[][pg. 39]{borusyakhull2021a} note,
\begin{quotation}
  \em
  The key challenge of applying our framework, absent true randomization, is in specifying plausible shock counterfactuals.
\end{quotation}
This challenge raises a basic question: how sensitive are causal conclusions to the researcher's specification of the shock assignment process?
The method we develop enables researchers to evaluate the sensitivity of their estimates to small or large deviations away from an assumed baseline distribution of shocks.
The method can be reliably implemented at scale with linear programming techniques through our companion \texttt{R} package \texttt{formulaiv}.

We use our method to reanalyze the empirical applications in both \cite{borusyakhull2023e} and \cite{borusyakhull2026e}.

\cite{borusyakhull2023e} analyze the effect of market access on employment using the roll-out of the high-speed rail system in China.
The authors find that a naive OLS estimate yields large positive effects, while their formula instrument approach produces a small positive effect that is indistinguishable from zero.
Our sensitivity analysis shows that small changes in the distribution of shocks used in their formula instrument lead to instrumental variable estimates that are anywhere from large negative effects to large positive effects.
We show that the specification test proposed by \cite{borusyakhull2023e} is unable to reject the null hypothesis that any of these alternative distributions are correctly specified.

\cite{borusyakhull2026e} analyze the effect of the Medicaid expansions on the take-up of private insurance.
The authors show that using a formula instrument allows one to tighten the precision on a simulated instrument approach \citep[similar to][]{freangrubersommers2017johe} by focusing attention on the population that is potentially affected by the reform.
However, doing so requires taking a stance on the distribution of Medicaid expansion shocks across states.
The authors do so by assuming that the ex-ante probability of expansion only depends on the party of the governor, so that, for example, Republican-led states like Michigan (did expand) and Alabama (did not) had equal probabilities of expanding, while Democrat-led states like Delaware (did expand) and Missouri (did not) also had equal ex-ante probabilities of expanding.
Our sensitivity analysis shows that changing the Republican-led probabilities to be non-homogeneous allows for formula instrument estimates that are consistent with a broad range of possible effects of both Medicaid eligibility and take-up.
The implication is that the reduction in variance obtained by \cite{borusyakhull2026e} comes with the risk of substantial bias from misspecification of the distribution of counterfactual shocks.

Our paper is relevant for a growing empirical literature that applies the \cite{borusyakhull2023e} formula instrument method.
Examples include \cite{dellolken2020res}, \cite{bosshartweigand2025}, \cite{buhlerdickens2025}, and \cite{moroninicolettisalvanestominey2025}.
Our results suggest that formula instrument approaches can be sensitive to the parametric assumption about the shock distribution.
Our method provides researchers an easy way to assess this sensitivity in their applications.

Our paper is also related to an old but growing theoretical literature on sensitivity analysis in statistics and econometrics.
More recent examples include \cite{conleyhansenrossi2010roeas}, \cite{nevorosen2012roeas}, and \cite{klinesantos2013qe}; see \cite{mastenpoirier2025wp} for a survey with an emphasis on linear models.
More related to our contribution is a smaller literature focused on sensitivity to parametric distributional assumptions in nonlinear models, for example \cite{chentamertorgovitsky2011cfdp1}, \cite{bonhommeweidner2022q}, \cite{christensenconnault2023e}, and \cite{gurussell2024wp}, although all of these authors consider settings much different than formula instruments.

The structure of the paper is as follows.
In Section \ref{sec-model}, we explain the formula instrument approach and the recentered IV estimator that comes out of it.
In Section \ref{sec-sensitivity}, we develop our method for sensitivity analysis.
In Section \ref{sec-china}, we use our method to reanalyze the application to market access in \cite{borusyakhull2023e}.
In Section \ref{sec-medicaid}, we use our method to reanalyze the application to Medicaid expansion in \cite{borusyakhull2026e}.
Section \ref{sec-conclusion} provides some brief concluding remarks.

\section{Formula instruments and the recentered IV estimator}
\label{sec-model}

We briefly review the methodology developed by \cite{borusyakhull2023e}.

The authors consider the linear model
\begin{align}
  y_{i} = \beta x_{i} + \varepsilon_{i},
  \label{eq-outcome-equation}
\end{align}
where $i$ indexes the unit for $i = 1,\ldots,N$, $y_{i}$ is an outcome, $x_{i}$ is an endogenous treatment variable, and $\varepsilon_{i}$ is a latent residual.
Both $y_{i}$ and $x_{i}$ are scalar and assumed to be sample mean zero for simplicity.
The authors assume access to an instrumental variable $z_{i}$.
Their methodology is also applicable to the case when $z_{i} = x_{i}$, which is the case they analyze in the application we revisit in Section \ref{sec-china}.

The authors assume that each $z_{i}$ is determined as a known function (or \emph{formula}) of two types of observable variables: a vector of exogenous shocks, $g \equiv (g_{1},\ldots,g_{K})$, and a vector of predetermined covariates, $w_{i}$.
To allow for spillovers, each $z_{i}$ can in general be determined by the collection of covariates $w \equiv (w_{1},\ldots,w_{N})$ from other units.
The formula is a function $f_{i}$---possibly depending on $i$---that maps $g$ and $w$ to $z_{i}$:
\begin{align}
  \label{eq-instrument-formula}
  z_{i} = f_{i}(g; w).
\end{align}
The function $f_{i}$ is assumed to be known for all $i$.
The shocks are assumed to be exogenous in the sense of being conditionally independent of the entire vector of latent residuals $\varepsilon \equiv (\varepsilon_{1},\ldots,\varepsilon_{N})$ \citep[][Assumption 1]{borusyakhull2023e}.

\begin{assumption}
  \label{as-shock-exogeneity}
  \textbf{(Shock Exogeneity)}
  $g \independent \varepsilon \vert w$, where $\independent$ denotes independence.
\end{assumption}

Assumption \ref{as-shock-exogeneity} implies that $z_{i}$ is independent of $\varepsilon_{i}$, conditional on $w_{i}$, but not unconditionally.
As a consequence, Assumption \ref{as-shock-exogeneity} is not sufficient to ensure that the linear IV estimator that uses $z_{i}$ as an instrument for $x_{i}$ will be consistent for $\beta$.
To see this, write \eqref{eq-outcome-equation} as
\begin{align}
  y_{i}
  =
  \beta x_{i}
  +
  \Exp[\varepsilon_{i} \vert g, w]
  +
  \left(
    \varepsilon_{i} - \Exp[\varepsilon_{i} \vert g, w]
  \right)
  =
  \beta x_{i}
  +
  \underbrace{
    \Exp[\varepsilon_{i} \vert w]
    +
    \eta_{i}
  }_{= \varepsilon_{i}},
  \label{eq-why-not-unconditionally-exogenous}
\end{align}
where the second equality invokes Assumption \ref{as-shock-exogeneity} and defines $\eta_{i} \equiv \varepsilon_{i} - \Exp[\varepsilon_{i} \vert g, w]$.
The new residual, $\eta_{i}$, satisfies $\Exp[\eta_{i} \vert z] = 0$ because of Assumption \ref{as-shock-exogeneity} and the formula relationship \eqref{eq-instrument-formula}:
\begin{align}
  \Exp[\eta_{i} \vert z]
  =
  \Exp\left[\Exp[\varepsilon_{i} \vert g,w,z] - \Exp[\varepsilon_{i} \vert g,w] \vert z\right]
  =
  \Exp\left[\Exp[\varepsilon_{i} \vert g,w] - \Exp[\varepsilon_{i} \vert g,w] \vert z\right]
  =
  0.
\end{align}
However, if $\Exp[\varepsilon_{i} \vert w]$ is a non-constant function of $w$, then $z_{i}$, which is also a function of $w$ via the formula \eqref{eq-instrument-formula}, will generally be correlated with the original residual, $\varepsilon_{i}$.

The traditional solution to this problem is to control for $w$.
In the context of \eqref{eq-why-not-unconditionally-exogenous}, this means specifying a functional form for $\Exp[\varepsilon_{i} \vert w]$.
For example, if $\Exp[\varepsilon_{i} \vert w] = w_{i}'\alpha$, then both $\beta$ and $\alpha$ can be consistently estimated by the linear IV estimator that uses $z_{i}$ as an instrument for $x_{i}$ while controlling for $w_{i}$, assuming sufficient independent variation in $z_{i}$.
The motivation for using the \cite{borusyakhull2023e} approach of leveraging the formula \eqref{eq-instrument-formula} is that specifying the correct functional form for $\Exp[\varepsilon_{i} \vert w]$ may be difficult, especially when $w$ is a complex set of controls.
\citet[][pp. 2161--2162]{borusyakhull2023e} argue in the context of three empirical examples that it would be challenging to choose the correct functional form to control for $w$.

The alternative proposed by \cite{borusyakhull2023e} is to instead model the conditional expectation of the instrument, $\mu_{i} \equiv \Exp[z_{i} \vert w]$.
If this conditional expectation were known, then the instrument could be recentered as $\tilde{z}_{i} \equiv z_{i} - \mu_{i}$.
While Assumption \ref{as-shock-exogeneity} is not sufficient to ensure that the original instrument, $z_{i}$, is uncorrelated with $\varepsilon_{i}$, it is sufficient to ensure that the recentered instrument, $\tilde{z}_{i}$, is uncorrelated with $\varepsilon_{i}$:
\begin{align}
  \Exp\left[\tilde{z}_{i}\varepsilon_{i}\right]
  =
  \Exp\left[\tilde{z}_{i}\Exp[\varepsilon_{i} \vert g,w]\right]
  =
  \Exp\left[\tilde{z}_{i}\Exp[\varepsilon_{i} \vert w]\right]
  =
  \Exp\left[\Exp[\tilde{z}_{i} \vert w]\Exp[\varepsilon_{i} \vert w]\right]
  =
  0.
\end{align}
This suggests using the linear IV estimator that instruments for $x_{i}$ with $\tilde{z}_{i}$ instead of $z_{i}$, which \cite{borusyakhull2023e} describe as the ``recentered IV'' (RIV):
\begin{align}
  \betariv
  \equiv
  \frac{\sum_{i=1}^{N}y_{i}\tilde{z}_{i}}{\sum_{i=1}^{N}x_{i}\tilde{z}_{i}}.
  \label{eq:riv}
\end{align}
Under the usual statistical conditions, $\betariv$ will be a consistent estimator of $\beta$.
Earlier examples of this argument can be found in the literature on partially linear models, notably \cite{robinson1988e}, ideas from which feature prominently in the modern literature on using machine learning to control for covariates in IV regressions \citep[for example,][Section 4.2]{chernozhukovchetverikovdemirerduflohansenetal2018ej}, and have also been used in the literature on marginal treatment effects \citep[for example,][]{carneiroheckmanvytlacil2011aer, andresen2018tsj}.

The benefit of using $\betariv$ is that there is no need to specify the functional form of $\Exp[\varepsilon_{i} \vert w]$.
The appeal of recentering the instrument turns on the relative difficulty of modeling $\mu_{i} \equiv \Exp[z_{i} \vert w]$ and $\Exp[\varepsilon_{i} \vert w]$.
Both are potentially complicated functions when $w$ is a complex vector of covariates.
The novel proposal of \cite{borusyakhull2023e} is that one can model $\mu_{i}$ by combining the formula \eqref{eq-instrument-formula} with the assumption that the conditional distribution of the shocks $g$, denoted $G(\cdot \vert w)$, is known by the researcher.
This requires maintaining the following assumption \citep[][Assumption 2]{borusyakhull2023e}, which the authors describe as a ``Known Assignment Process''.

\begin{assumption}
  \label{as-known-assignment-process}
  \textbf{(Known Assignment Process)}
  The distribution of $g$ conditional on $w$ is known and given by $G(g \vert w)$ for all supported $w$.
\end{assumption}

Assumption \ref{as-known-assignment-process} and the formula \eqref{eq-instrument-formula} enable direct computation of $\mu_{i}$ through simulation.
For example, \cite{borusyakhull2023e} suggest choosing $G(\cdot \vert w) = G(\cdot)$ to be the uniform distribution over the set of all permutations of the observed realization of $g \equiv (g_{1},\ldots,g_{K})$, independently of $w$.
There are $K!$ permutations of the $K$ elements of $g$, so this suggestion implies the assumption that $G(\cdot \vert w)$ places equal mass $1/K!$ on each permutation formed from the components of the realized $g$.
When $K!$ is a large number, the authors propose approximating $\mu_{i}$ with a subset of $S$ permutations.
With $\mu_{i}$ (or a sufficient approximation) in hand, the recentered IV estimator $\betariv$ in \eqref{eq:riv} can then be constructed by using $\tilde{z}_{i} \equiv z_{i} - \mu_{i}$ as an instrument for $x_{i}$, without controlling for covariates.

\section{Sensitivity to the known assignment process}
\label{sec-sensitivity}

In this section, we develop a systematic sensitivity analysis that relaxes Assumption \ref{as-known-assignment-process}.

Our object of interest is the joint distribution $G(\cdot \vert w)$ of the shock vector $g \equiv (g_{1},\ldots,g_{K})$.
We assume for simplicity that $G(\cdot \vert w) = G(\cdot)$ does not depend on $w$, since this is the case in both of the applications we consider; however, this is not essential to what follows.
We represent $G$ through a finite support $\{(g_{1s},\ldots,g_{Ks})\}_{s=1}^{S}$ of shock realizations, together with a vector of probabilities $p \equiv (p_{1},\ldots,p_{S})$ assigned to them, where $p_{s} \equiv \Prob_{G}[g = (g_{1s},\ldots,g_{Ks})]$.\footnote{
  Our analysis can be extended to cases where $G$ has a continuous distribution; see Appendix \ref{sec-general-assignment}.
}
Then
\begin{align}
  \mu_{i}
  =
  \sum_{s=1}^{S}f_{i}((g_{1s},\ldots,g_{Ks}), w)\Prob_{G}[g = (g_{1s},\ldots,g_{Ks})]
  \equiv
  \sum_{s=1}^{S}f_{is}p_{s},
  \label{eq-mu-as-discrete-sum}
\end{align}
where $f_{is} \equiv f_{i}((g_{1s},\ldots,g_{Ks}), w)$.
The vector $p$ must live in the $S$-dimensional simplex of non-negative numbers that sum to one, which we denote by $\Delta^{S}$.

We consider sensitivity of the recentered IV estimate to the choice of $p$ as it deviates from some baseline $\bar{p}$ across some pre-determined set $\mathcal{P} \subseteq \Delta^{S}$.
For example, $\bar{p}$ might be the uniform distribution used by \cite{borusyakhull2023e}, which has $\bar{p}_{s} = 1/S$ for all $s$.
We consider two ways of specifying the sensitivity set $\mathcal{P}$, intended to capture different ways of measuring deviations between $p$ and $\bar{p}$.

The first way is to require each component of $p$ to be within $\kappa \geq 1$ multiples of its corresponding component of $\bar{p}$ by restricting $p$ to the set
\begin{align}
  \PJ(\kappa \vert \bar{p})
  \equiv
  \left\{
    p \in \Delta^{S}:
    \frac{1}{\kappa}
    \bar{p}_{s}
    \leq
    p_{s}
    \leq
    \kappa
    \bar{p}_{s}
    \quad
    \text{for all $s = 1,\ldots,S$}
  \right\}.
  \label{eq-main-sensitivity-set}
\end{align}
Setting $\kappa = 1$ makes $\PJ(1 \vert \bar{p}) = \{\bar{p}\}$ a singleton, while as $\kappa \rightarrow \infty$, the set $\PJ(\kappa \vert \bar{p})$ becomes closer to the simplex, $\Delta^{S}$.
We call $\PJ(\kappa \vert \bar{p})$ the \emph{joint sensitivity set} because it measures deviations from $\bar{p}$ in terms of the entire joint distribution $p$.
This measure is used in the robust Bayes literature where $\bar{p}$ is viewed as a baseline prior \citep[][]{lavine1991jotasa, wassermankadane1992jotasa}.

The second way is to constrain the marginal distributions of each $g_{k}$ rather than the entire joint distribution.
For a joint distribution $p \in \Delta^{S}$, the implied marginal probability that $g_{k} = h$ is
\begin{align}
  q_{k}(h \vert p)
  \equiv
  \sum_{s=1}^{S}\IndicSmall{g_{ks} = h}p_{s}.
  \label{eq:marginal-from-joint}
\end{align}
Let $\bar{q} = (\bar{q}_{1},\ldots,\bar{q}_{K}) \equiv (q_{1}(\cdot \vert \bar{p}),\ldots,q_{K}(\cdot \vert \bar{p}))$ denote the baseline collection of marginals produced from the baseline joint distribution $\bar{p}$ via \eqref{eq:marginal-from-joint}.
We define the \emph{marginal sensitivity set} to be the set of joint distributions whose implied marginals are within $\delta \geq 1$ multiples of $\bar{q}$,
\begin{align*}
  \PM(\delta \vert \bar{q})
  \equiv
  \Bigg\{
    p \in \Delta^{S} :
    \frac{1}{\delta}
    \bar{q}_{k}(h)
    \leq
    &q_{k}(h \vert p)
    \leq
    \delta
    \bar{q}_{k}(h) \\
    &\quad
    \text{for each $h \in \mathcal{H}_{k}$ and $k = 1,\ldots,K$}
  \Bigg\}.
\end{align*}
For example, in the application in Section \ref{sec-china-replication}, each shock $g_{k}$ is a binary event, so setting $\mathcal{H}_{k} = \{1\}$ for each $k$ makes $\PM(\delta \vert \bar{q})$ the set of $p$ whose event probabilities for each shock $k$ are within $\delta$ multiples of the baseline event probabilities $\bar{q}_{k}(1)$.
Note that unlike the joint sensitivity set, the marginal sensitivity set does not collapse to a singleton at $\delta = 1$, because many joint distributions can share the same marginals.

Each choice of $p \in \mathcal{P}$ produces a different recentered IV estimator by changing $\mu_{i}$ in \eqref{eq-mu-as-discrete-sum}.
We denote this dependence by writing $\mu_{i}(p)$.
The recentered IV using $p$ is then $\tilde{z}_{i}(p) \equiv z_{i} - \mu_{i}(p)$ and the recentered IV estimator is
\begin{align}
  \betariv(p)
  \equiv
  \frac{
    \sum_{i=1}^{N}y_{i}\tilde{z}_{i}(p)
  }{
    \sum_{i=1}^{N}x_{i}\tilde{z}_{i}(p)
  },
  \label{eq-pi-constraint-set-multiplicative}
\end{align}
noting again that $y_{i}$ and $x_{i}$ are assumed to have sample mean zero for simplicity.
The recentered IV estimator varies as $p$ ranges across a sensitivity set $\mathcal{P}$, such as $\PJ(\kappa \vert \bar{p})$ or $\PM(\delta \vert \bar{q})$.
To handle the possibility that $\betariv(p)$ is undefined because the denominator $D(p) \equiv\sum_{i=1}^{N}x_{i}\tilde{z}_{i}(p)$ of $\betariv(p)$ is zero, we
define the set
\begin{align*}
  \mathcal{P}_{D \neq 0}
  \equiv
  \left\{
    p \in \mathcal{P} :
    D(p)
    \neq
    0
  \right\}.
\end{align*}
Then the smallest and largest values that $\betariv(p)$ can take are
\begin{align}
  \label{eq-fractional-lp}
  \betalb(\mathcal{P})
  \equiv
  \inf_{p \in \mathcal{P}_{D \neq 0}}
  \betariv(p)
  \quad
  \text{and}
  \quad
  \betaub(\mathcal{P})
  \equiv
  \sup_{p \in \mathcal{P}_{D \neq 0}}
  \betariv(p).
\end{align}
The following proposition shows that these extremal values can be found by solving linear programs as long as $\mathcal{P}$ is a polyhedron, such as $\PJ(\kappa \vert \bar{p})$ or $\PM(\delta \vert \bar{q})$.

\begin{proposition}
  \label{prop-sensitivity}
  Suppose that $\mathcal{P} \subseteq \Delta^{S}$ is a polyhedron, written as $\mathcal{P} = \{p \in \Delta^{S} : Ap \leq c\}$ for some known matrix $A$ and vector $c$.
  If $D(p) \geq 0$ for all $p \in \mathcal{P}$ then
  \begin{align}
    \betalb(\mathcal{P})
    =
    \min_{
      \phi \in \re^{S}, \tau \in \re
    }
    \quad
    &\sum_{i=1}^{N}
    y_{i}z_{i}\tau
    -
    \sum_{i=1}^{N}
    \sum_{s=1}^{S}
    y_{i}f_{is}\phi_{s} \notag \\
    \text{s.t.}
    \quad
    &\tau \geq 0, \phi_{s} \geq 0 \text{ for all $s = 1,\ldots,S$} \notag \\
    &\sum_{s=1}^{S} \phi_{s} = \tau \notag \\
    &A\phi \leq c\tau \notag \\
    \quad
    &\sum_{i=1}^{N}x_{i}z_{i}\tau
    -
    \sum_{i=1}^{N}
    \sum_{s=1}^{S}
    x_{i}f_{is}\phi_{s}
    =
    1,
    \label{eq-sensitivity-lp}
  \end{align}
  and $\betaub(\mathcal{P})$ is given by the corresponding maximization problem.\footnote{
    We use the usual convention here of setting $\betalb(\mathcal{P}) = -\infty$ if the minimization problem is unbounded and $\betaub(\mathcal{P}) = +\infty$ if the maximization problem is unbounded.
  }
  Moreover, for any real number $b \in [\betalb(\mathcal{P}), \betaub(\mathcal{P})]$, there exists a $p \in \mathcal{P}$ such that $\betariv(p) = b$.
  If instead $D(p) \leq 0$ for all $p \in \mathcal{P}$ then the same statement is true after two changes: (i) change the last constraint in \eqref{eq-sensitivity-lp} from $1$ to $-1$ and (ii) take $-\betaub(\mathcal{P})$ to be the optimal value of the minimization problem and take $-\betalb(\mathcal{P})$ to be the optimal value of the corresponding maximization problem.
  If $D(p)$ takes both positive and negative values as $p$ ranges over $\mathcal{P}$, and if it is not the case that $\betariv(p)$ is constant for all $p \in \mathcal{P}_{D \neq 0}$, then $\betalb(\mathcal{P}) = -\infty$ and $\betaub(\mathcal{P}) = +\infty$.
\end{proposition}

Proposition \ref{prop-sensitivity} provides a computationally tractable way to compute all of the possible values that the recentered IV estimator $\betariv(p)$ can take as $p$ varies across the sensitivity set $\mathcal{P}$.
The linear programs have $S + 1$ variables and a similar number of constraints if $\mathcal{P}$ is taken to be the joint or marginal sensitivity set.
This makes the programs straightforward to solve even if $S$ is quite large.
The justification of Proposition \ref{prop-sensitivity} recognizes that $\betalb(\mathcal{P})$ and $\betaub(\mathcal{P})$ are the optimal values of linear fractional programs because $\betariv(p)$ is the ratio of two affine functions of $p$.
Applying the \cite{charnescooper1962nrlq} transformation to \eqref{eq-fractional-lp} yields the linear program \eqref{eq-sensitivity-lp}; see, for example, \citet[][pg. 151]{boydvandenberghe2004}.

\section{Reevaluating the effects of market access in China}
\label{sec-china}

In this section, we use Proposition \ref{prop-sensitivity} to reanalyze the application in \cite{borusyakhull2023e}.

\subsection{Replication}
\label{sec-china-replication}

\cite{borusyakhull2023e} use a two-period panel of 275 subprovince-level administrative divisions (``prefectures'') in mainland China.
Regional employment in prefecture $i$ is defined as urban employment as taken from the Chinese City Statistical Yearbooks.
Market access in prefecture $i$, year $t$, is defined as
\begin{align}
  \label{eq:ma-formula}
  \ma_{it} =\sum_{j=1}^{N} \exp \left(-0.02 \tau_{ijt}\right) \times \pop_{j,2000}, \quad
  \text{for $t = 2007, 2016$,}
\end{align}
where $\pop_{j, 2000}$ is the population of prefecture $j$ in $2000$, and $\tau_{ijt}$ is the predicted travel time between prefectures $i$ and $j$ in year $t$.

The travel time $\tau_{ijt}$ is determined in part by the presence of high-speed rail (HSR) connections between the prefectures.
\cite{borusyakhull2023e} compute $\tau_{ijt}$ using comprehensive data on the evolution of the Chinese HSR network.
The network includes 150 potential total lines: 83 lines that opened between 2007 and 2016, 66 additional lines that were planned or under construction by April 2019, but had not yet opened by the end of 2016, as well as one line between Qinhuangdao and Shenyang that opened in 2003.

This roll-out of HSR lines produces variation in MA over time.
The authors define their endogenous variable $x_{i}$ as this change over the course of their two-period panel: $x_{i} \equiv \log \ma_{i,2016} - \log \ma_{i,2007}$.
They take the outcome $y_{i}$ to be the corresponding change in urban employment between 2007 and 2016.
The empirical challenge is to determine the causal effect of $x_{i}$ on $y_{i}$.
As \cite{borusyakhull2023e} discuss, this is difficult because $x_{i}$ is correlated with geography $w_{i}$, which may be correlated with unobserved determinants of employment growth, such as local productivity shocks.

\cite{borusyakhull2023e} apply the recentered IV approach to this problem.
The shock sequence $g \equiv (g_{1},\ldots,g_{K})$ is a vector of $K = 150$ binary shocks for each HSR line $k$, with $g_{k} = 1$ denoting that a line opened by 2016 and $g_{k} = 0$ denoting that it did not open.
Assumption \ref{as-shock-exogeneity} requires these line openings to be independent of unobserved determinants of employment growth, perhaps conditional on geographic controls $w_{i}$.
To operationalize Assumption \ref{as-known-assignment-process}, \cite{borusyakhull2023e} assume that $G$ is a uniform distribution over a fixed support of $S = 1999$ draws of $g$.\footnote{
  The authors need to do this because the formula $f$ implied by the market access function \eqref{eq:ma-formula} is non-separable across $g_{k}$ through their interdependence in $\tau_{ijt}$.
  In Section \ref{sec-medicaid}, we consider an example where the support of $g$ is not constrained in this way.
}
For each draw, $(g_{1s},\ldots,g_{Ks})$, they recompute the travel time variable $\tau_{ijt}$, then construct $\mu_{i} \equiv \Exp[x_{i} \vert w_{i}] = S^{-1}\sum_{s=1}^{S}f_{i}((g_{1s},\ldots,g_{Ks}), w)$ using the formula for $x_{i} \equiv \log \ma_{i,2016} - \log \ma_{i,2007}$ implied by \eqref{eq:ma-formula}.
Note that this application has $z_{i} = x_{i}$, which is a special case of the formula IV framework that might be more appropriately called ``formula OLS.''

We are able to replicate the results in \cite{borusyakhull2023e} exactly by using the same sample of $S$ permuted $g$ vectors, which the authors included in their replication package.
We briefly review these results, which are the same as in Table I of \cite{borusyakhull2023e}. An unadjusted OLS estimate of $y_{i}$ on $x_{i}$ yields a statistically significant estimate of $.232$, which would be interpreted as an elasticity of employment with respect to the market access measure.
Controlling for geographic measures lowers this to $.133$, which is still statistically significant (standard error $.064$).
By contrast, the authors' recentered IV estimate with no covariates produces a statistically insignificant point estimate of $.084$ with a similar standard error of $.097$.
Controlling for covariates lowers the recentered IV estimate to $.056$, with a standard error of $.089$.

\subsection{Sensitivity to the assumed assignment process}
\label{sec-china-sensitivity}

In the notation of Section \ref{sec-sensitivity}, the shock distribution $\pbh$ used by \cite{borusyakhull2023e} amounts to setting the probability of each of the $s=1,\ldots,1999 \equiv S$ drawn simulations to be $\bar{p}_{\textsc{bh},s} \equiv 1/1999 \approx .0005$.
Their reported estimate is $\betariv(\pbh)$.
Figure \ref{fig:est_joint_no_controls} shows how sensitive $\betariv(\pbh)$ is to this choice of $\pbh$, with sensitivity measured in terms of the joint sensitivity set $\PJ(\kappa \vert \pbh)$ and its parameter $\kappa$.
For example, a value of $\kappa = 5$ on the x-axis allows for a distribution of counterfactual network configurations with $p_{s}$ between $[1/(5 \times 1999), 5/1999] \approx [.0001, .0025]$ for each $s$, while still requiring $\sum_{s=1}^{S}p_{s} = 1$.
The y-axis of Figure \ref{fig:est_joint_no_controls} shows the set $[\betalb(\PJ(\kappa \vert \pbh)), \betaub(\PJ(\kappa \vert \pbh))]$, which contains all values of $\betariv(p)$ that one could obtain for a $p \in \PJ(\kappa \vert \pbh)$.

\fig%
{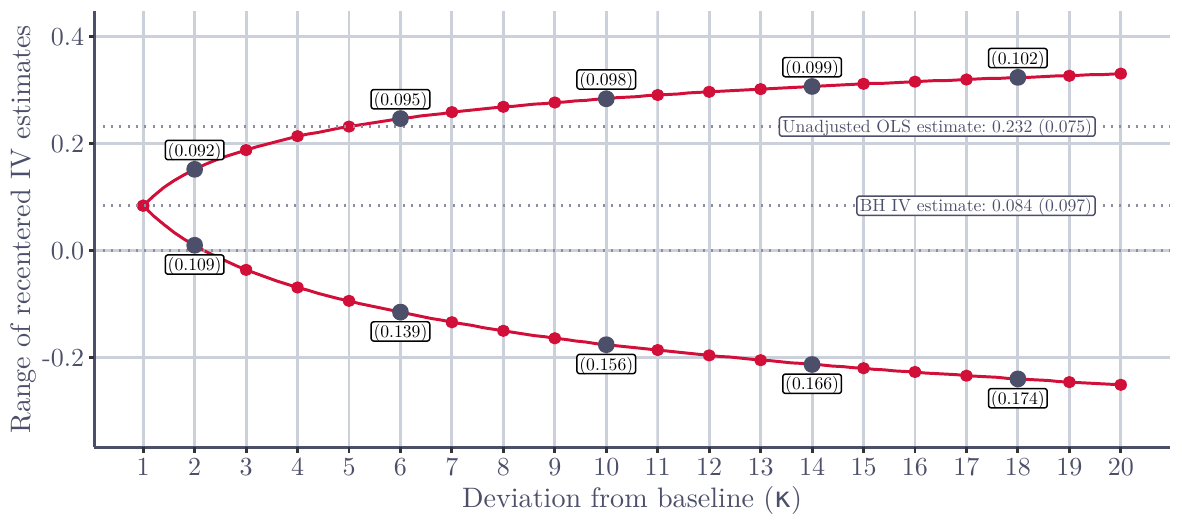}%
{fig:est_joint_no_controls}%
{Sensitivity to assumed joint distribution in \cite{borusyakhull2023e}}%
{%
  Bounds from solving \eqref{eq-sensitivity-lp} with $\mathcal{P} = \PJ(\kappa \vert \pbh)$ for different values of $\kappa$.
  The horizontal lines show the OLS and recentered IV estimates from \citet[][Table I, Panel A, Column 2]{borusyakhull2023e}.
  Spatially-clustered \cite{conley1999joe} standard errors are shown in parentheses, following the same specification as in \cite{borusyakhull2023e}.
  For the bounds, these show the standard errors at the optimizer for the program.
}

At $\kappa = 1$, the bounds collapse to the baseline estimate reported by \cite{borusyakhull2023e}.
As $\kappa$ increases, the bounds widen, reflecting ambiguity in the specification of Assumption \ref{as-known-assignment-process}.
For example, with $\kappa = 5$, the set of recentered IV estimates one can obtain includes everything from substantial negative employment effects of about $-.100$ to substantial positive employment effects that are about equal to the unadjusted OLS estimate of $.232$.
The implication is that changes in the assumed shock distribution (Assumption \ref{as-known-assignment-process}) can lead the recentered IV estimate to be as potentially misleading about positive employment effects as the uncontrolled OLS estimate, while also leaving open the possibility of \emph{negative} employment effects.
Figure \ref{fig:est_joint_with_controls} shows that controlling for geographic covariates leads to similar conclusions.

Is $\kappa = 5$ large or small?
The baseline choice of $\bar{p}_{\textsc{bh},s} = 1/1999 \approx .0005$ made by \cite{borusyakhull2023e} requires each of the $1999$ shocks to have an equal probability that is small, with no single shock realization occurring in more than $.05\%$ of potential draws of the underlying data generating process.
Setting $\kappa = 5$ means that none of the $1999$ shock configurations can occur in more than $.25\%$ or less than $.01\%$ of these draws.
It is not clear how one could reason about the exact magnitude of so many small probabilities simultaneously, suggesting that $\kappa = 5$ is rather small compared to the baseline of $\kappa = 1$.
As Figure \ref{fig:est_joint_no_controls} shows, increasing $\kappa$ to $10$ leads to even greater ambiguity, while still imposing the mild restriction that no possible shock realization occurs in more than $.5\%$ of draws.

\fig%
{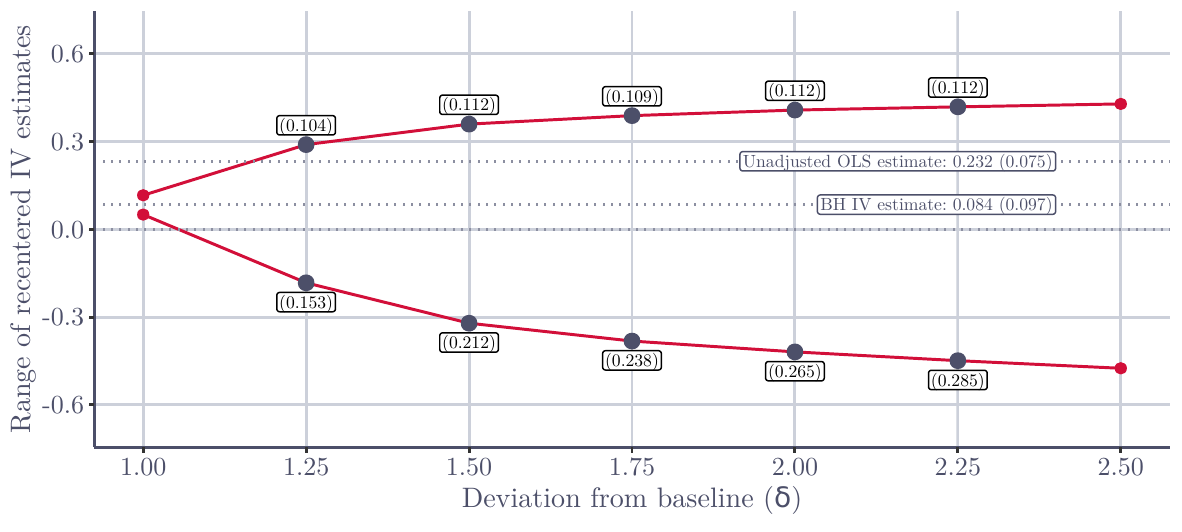}%
{fig:est_marginal_no_controls}%
{Sensitivity to assumed marginal distribution in \cite{borusyakhull2023e}}%
{%
  Bounds from solving \eqref{eq-sensitivity-lp} with $\mathcal{P} = \PM(\delta \vert \qbh)$ for different values of $\delta$.
  See notes for Figure \ref{fig:est_joint_no_controls}.
}

Figure \ref{fig:est_marginal_no_controls} shows sensitivity measured across the marginal set $\PM(\delta \vert \qbh)$ with $\mathcal{H}_{k} = \{1\}$ for each $k$.
The probability of each shock $g_{k}$ being equal to one represents the probability that HSR line $k$ opened by 2016.
The joint distribution $\pbh$ used by \cite{borusyakhull2023e} implies marginal probabilities $\qbh$ that have most line opening probabilities between roughly $.4$ and $.6$, with a few lines pegged to an opening probability of one.
For a line with a probability of $.5$, setting $\delta = 1.25$ means that $\PM(\delta \vert \qbh)$ contains joint distributions $p$ that admit marginal (line opening) probabilities between $.4$ and $.625$.
As Figure \ref{fig:est_marginal_no_controls} shows, even this mild relaxation is consistent with a broad range of recentered IV estimates that produce anything from large negative to large positive estimates.

\subsection{Specification tests}
\label{sec-china-specification-test}

\citet[][Section 3.5]{borusyakhull2023e} suggest that Assumption \ref{as-known-assignment-process} can be tested using randomization inference with test statistic equal to the sample covariance between the recentered instrument and the implied residual.
They conduct this test for their market access application and report a p-value of $.711$, failing to reject the null that the known assignment process is correctly specified.
They interpret this result as ``validating'' their specification of the HSR assignment process \citep[][pg. 2174]{borusyakhull2023e}.

\fig%
{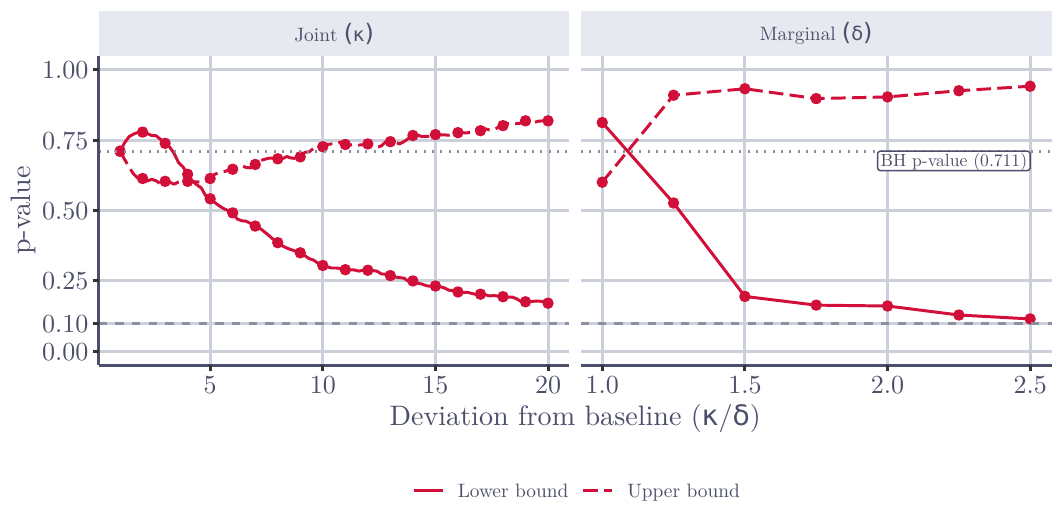}%
{fig:false_no_controls}%
{P-values from the \cite{borusyakhull2023e} specification test}%
{%
  P-values from the randomization inference test proposed by \cite{borusyakhull2023e}.
  The left-hand facet shows results with $\mathcal{P} = \PJ(\kappa \vert \pbh)$ and the right-hand facet shows results with $\mathcal{P} = \PM(\delta \vert \qbh)$.
  The dotted line is the same p-value reported by \citet[][Table II, Column 3]{borusyakhull2023e} for their baseline specification.
}

Figure \ref{fig:false_no_controls} shows p-values from the same test conducted for assignment processes that yield the lower and upper bounds for each $\kappa$ and $\delta$ considered in Figures \ref{fig:est_joint_no_controls} and \ref{fig:est_marginal_no_controls}.
The p-values do not cross even a conservative conventional threshold such as $.10$ for any value of $\kappa$ or $\delta$: the specification test never rejects.
This is despite the fact that we know that the assignment processes at the lower and upper bounds and at different values of $\kappa/\delta$ are inconsistent with one another.
The implication is that the test proposed by \cite{borusyakhull2023e} has low power for detecting violations of Assumption \ref{as-known-assignment-process}.

\subsection{Alternative baseline distributions}
\label{sec-china-baseline-distributions}

The results in Figures \ref{fig:est_joint_no_controls}--\ref{fig:false_no_controls} show that recentered IV estimates are sensitive to deviations from the baseline shock distribution $\pbh$ chosen by \cite{borusyakhull2023e} in a way that is not detectable through their specification test.
In this section, we examine whether $\pbh$ is a sensible starting point.

\fig%
{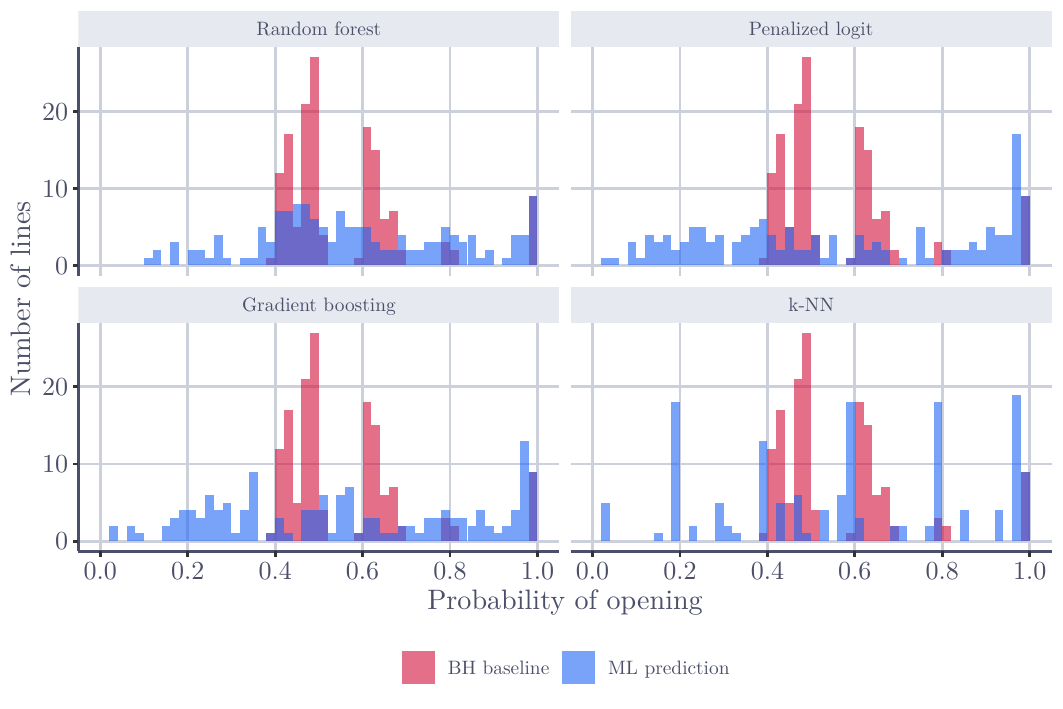}%
{fig:histogram_ml}%
{Marginal probability of a line opening}%
{%
  Each facet shows the out-of-sample histogram of marginal probabilities of opening across the 150 HSR lines for a machine learning model compared to the marginal probabilities $\qbh$ generated by the uniform shock distribution $\pbh$ used by \cite{borusyakhull2023e}.
  The specification and training of the models is discussed in Appendix \ref{sec-sa-predicting-marginal-shock-probs}.
}

While the choice of $\pbh$ specifies only one joint probability over the $S = 1999$ possible shock realizations, it implies $K = 150$ marginal probabilities for each of the HSR lines in the data.
This suggests a data-driven exercise: for each HSR line, we train machine learning algorithms that use the predetermined characteristics of the line in 2007 to predict whether the line would be opened by 2016.
We fit four learners: a random forest, penalized logistic regression, gradient-boosted trees, and $k$-nearest neighbors; Appendix \ref{sec-sa-predicting-marginal-shock-probs} contains details on how we specified and trained them.

Figure \ref{fig:model_eval} shows that---unsurprisingly---each of these learners provides better out-of-sample predictions than the implicit prediction $\qbh$ generated by the uniform shock distribution $\pbh$ used by \cite{borusyakhull2023e}.
Figure \ref{fig:histogram_ml} compares the histograms of line openings for the four models to $\qbh$.
Whereas $\qbh$ has many line opening probabilities concentrated around $.4$ and $.6$, the learners recognize that some lines were considerably more or less likely to open for reasons that could be predicted from their predetermined characteristics.
This provides additional evidence against the suggestion that the shocks should be viewed as ``exchangeable,'' a condition which \citet[][pg. 2166]{borusyakhull2023e} appeal to as a sufficient condition to support their choice of the uniform distribution $\pbh$.

\fig%
{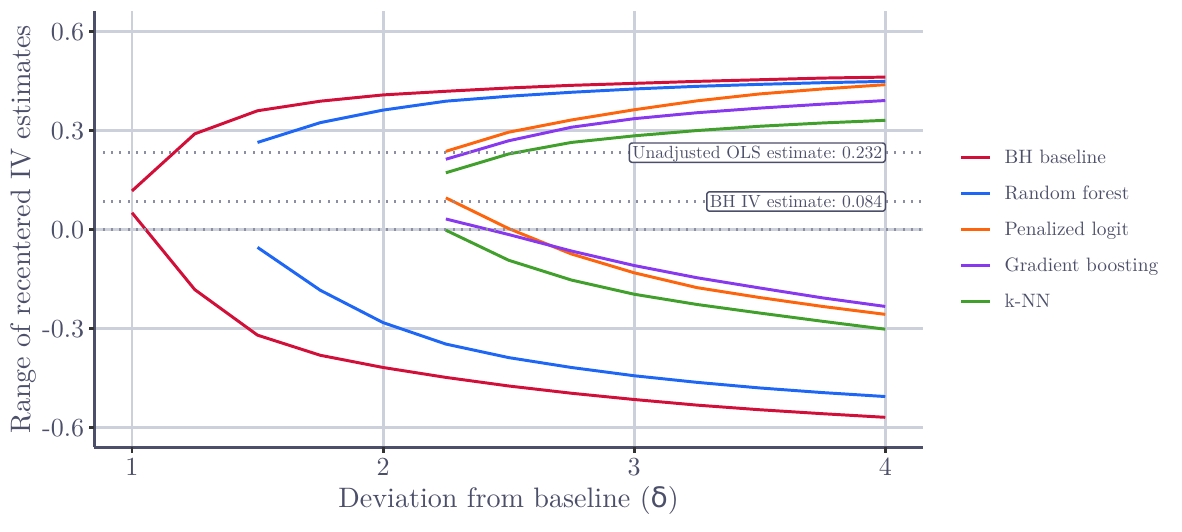}%
{fig:est_marginal_ml_no_controls}%
{Sensitivity to machine learning marginal distributions}%
{%
  Bounds from solving \eqref{eq-sensitivity-lp} with $\mathcal{P} = \PM(\delta \vert \bar{q}_{m})$ for different values of $\delta$ and the $\bar{q}_{1},\ldots,\bar{q}_{4}$ generated by the four learners shown in Figure \ref{fig:histogram_ml}.
}

Figure \ref{fig:est_marginal_ml_no_controls} reports a sensitivity analysis comparable to Figure \ref{fig:est_marginal_no_controls} when $\PM(\delta \vert \bar{q}_{m})$ is specified relative to the marginal distributions produced by the four machine learning models, $\bar{q}_{1},\ldots,\bar{q}_{4}$.
We start the x-axis for each model at the first value of $\delta$ for which it's possible to find any valid probability distribution $p \in \PM(\delta \vert \bar{q}_{m})$ that rationalizes $\bar{q}_{m}$.
For the best-performing model, this requires taking $\delta$ past two, implying a relaxation of $[.25, 1.00]$ for a line-opening probability of $.5$.
This suggests that the support of $S = 1999$ counterfactual shocks used by \cite{borusyakhull2023e} is itself hard to rationalize with the data.
Even so, deviations around each of the baselines provided by the machine learning models show the same type of sensitivity as for the baseline used by \cite{borusyakhull2023e}.
This suggests that the sensitivity found in Figures \ref{fig:est_joint_no_controls} and \ref{fig:est_marginal_no_controls} is a consequence of the formula IV idea itself, rather than the specific choice of baseline reference distribution.

\section{Reevaluating the impacts of Medicaid expansion}
\label{sec-medicaid}

\cite{borusyakhull2026e} apply the formula instrument idea to evaluate the impact of Medicaid eligibility on private insurance take-up using the partial state-level expansion of Medicaid that occurred under the Affordable Care Act (ACA) in 2014.
The authors use a repeated cross-section of individuals $i$ from the American Community Survey (ACS).
The endogenous variable $x_{i} \in \{0,1\}$ is individual $i$'s eligibility for Medicaid and the outcome $y_{i}$ is a measure of private insurance take-up.
The authors propose a linear model of the form
\begin{align}
  \label{eq:medicaid-outcome}
  y_{i} = \beta x_{i} + \alpha_{s(i)} + \tau_{r_{s(i)}, t(i)} + \varepsilon_{i},
\end{align}
where $s(i)$ and $t(i)$ denote the state and year of individual $i$, $\alpha_{s(i)}$ are state fixed effects, $r_{k}$ is an indicator for whether the government of state $k$ in 2013 is a Republican, $\tau_{r, t}$ are party-by-year fixed effects, and $\varepsilon_{i}$ is an unobservable.
The concern is that $x_{i}$ and $\varepsilon_{i}$ may be correlated through the individual characteristics $c_{i}$.

The authors use the binary expansion decisions $g_{k} \in \{0,1\}$ for each state $k$ as the ``shocks.''
One way to do this is to instrument for $x_{i}$ using $z_{i} \equiv g_{s(i)}\IndicSmall{t(i) = \text{2014}}$, which produces an instrumented difference-in-differences estimate of $\beta$.
The authors describe this as a simulated instrument along the lines of \cite{curriegruber1996tqjoe} or \cite{freangrubersommers2017johe}.\footnote{
  The simulated instrument terminology may be misleading here because $z_{i}$ is binary, so lacks any variation intensity across states.
  We are following \cite{borusyakhull2023e} in our usage of the phrase.
}
A drawback of this approach is that many individuals have no variation in $x_{i}$ regardless of the value of $z_{i}$, for example if they are ineligible for Medicaid either with or without the expansion.
This dilutes the relevance of $z_{i}$ for $x_{i}$, making estimates of $\beta$ relatively imprecise.

\cite{borusyakhull2026e} propose a formula instrument alternative based on knowledge of how Medicaid eligibility is determined:
\begin{align}
  \label{eq:medicaid-treatment}
  x_{i} =
  h^{t(i)}
  (c_{i}, e_{s(i)}^{\text{2013}}, g_{s(i)}, e^{\Delta}_{s(i)}),
\end{align}
where $h^{t(i)}$ is a known, year-specific function that determines Medicaid eligibility, $c_{i}$ are individual characteristics such as income, work status, or parental status, $e_{k}^{\text{2013}}$ is the Medicaid eligibility policy of state $k$ in 2013, and $e_{k}^{\Delta}$ includes other changes in 2014 to Medicaid coverage in state $k$.
They propose recentering the instrument $z_{i} \equiv h^{t(i)}(c_{i}, e_{s(i)}^{\text{2013}}, g_{s(i)}, \emptyset)$ that ignores the non-ACA eligibility changes $e_{k}^{\Delta}$.
Recentering this instrument via Assumption \ref{as-known-assignment-process} is necessary for it to be exogenous because $z_{i}$ depends on individual characteristics $c_{i}$ that are likely also reflected in $\varepsilon_{i}$.

The model that the authors propose for Assumption \ref{as-known-assignment-process} is based on the assumption that
\begin{align}
  \label{eq:expansion-probabilities}
  \Prob[g_{s(i)} = 1 \vert w_{i}]
  =
  \Prob[g_{s(i)} = 1 \vert r_{s(i)}]
  =
  \pi(0)(1-r_{s(i)}) + \pi(1)r_{s(i)}
  =
  \pi(r_{s(i)}),
\end{align}
where $w_{i}$ collects $c_{i}, s(i), t(i), e_{s(i)}^{\text{2013}}$, and $r_{s(i)}$.
That is, the probability that state $k$ expands is a constant function $\pi(r_{k})$ of whether its governor in 2013 was a Republican, $r_{k}$.
This implies that, for example, two Republican-led states like Michigan and Alabama---one of which expanded and one of which did not---had ex-ante equal probabilities of adopting the ACA expansion.
Given \eqref{eq:expansion-probabilities}, the conditional expectation of $z_{i}$ given $w_{i}$ is
\begin{align}
  \mu_{i}
  \equiv
  \Exp[z_{i} \vert w_{i}]
  &=
  h^{t(i)}(c_{i}, e_{s(i)}^{\text{2013}}, 0, \emptyset)
  +
  \Exp[g_{s(i)} \vert w_{i}]
  a_{i}
  =
  h^{t(i)}(c_{i}, e_{s(i)}^{\text{2013}}, 0, \emptyset)
  +
  \pi(r_{s(i)})
  a_{i},
  \label{eq:medicaid-instrument-mean}
\end{align}
where $a_{i}$ is a binary indicator for whether individual $i$'s eligibility would have been affected by an expansion in 2014:
\begin{align*}
  a_{i}
  \equiv
  \IndicSmall{h^{2014}(c_{i}, e_{s(i)}^{\text{2013}}, 0, \emptyset)
    \neq
  h^{2014}(c_{i}, e_{s(i)}^{\text{2013}}, 1, \emptyset)}.
\end{align*}
From \eqref{eq:medicaid-instrument-mean} we get that the recentered instrument $\tilde{z}_{i} \equiv z_{i} - \mu_{i}$ is
\begin{align*}
  \tilde{z}_{i}
  =
  \left(
    g_{s(i)} - \pi(r_{s(i)})
  \right)
  a_{i}.
\end{align*}

Unlike the simulated instrument, this recentered instrument $\tilde{z}_{i}$ is mechanically zero for individuals with $a_{i} = 0$, whose eligibility would have been unaffected by an expansion in their state.
Using $\tilde{z}_{i}$ as an instrument therefore numerically drops these individuals, raising the hope that the resulting recentered IV estimator may be more precise than the simulated IV estimator.
As a practical matter, it also means that the recentered IV estimator using $\tilde{z}_{i}$ is numerically equivalent to an IV estimator that uses $g_{s(i)} - \pi(r_{s(i)})$ as an instrument for $x_{i}$ among the subsample of affected individuals $a_{i} = 1$.
Because the authors already include party-by-year fixed effects $\tau_{r_{s(i)}, t(i)}$ in \eqref{eq:medicaid-outcome}, this in turn is equivalent to just using $g_{s(i)}\IndicSmall{t(i) = \text{2014}}$ as an instrument for $x_{i}$, the same as in the simulated instrument but now only among the subsample with $a_{i} = 1$.

The key to this equivalence is \eqref{eq:expansion-probabilities}, which requires all Republican-led states to have had the same ex-ante probability of expanding.
This assumption may be concerning to observers of U.S.\@ politics.
Without it, one would be unable to recenter the instrument without specifying the distribution over the expansion indicators $g \equiv (g_{1},\ldots,g_{43})$, as in the market access application in Section \ref{sec-china}.
If there is within-party heterogeneity in the expansion probability, then the expansion probabilities will no longer be absorbed by the party-by-year fixed effects used in \eqref{eq:medicaid-outcome}.

To evaluate sensitivity to \eqref{eq:expansion-probabilities}, we apply Proposition \ref{prop-sensitivity} to allow states with the same party to have different expansion probabilities.
We take $\{g_{1s},\ldots,g_{Ks}\}_{s=1}^{S}$ to be the full set of $S = 2^{43}$ possible binary realizations.\footnote{
  Because $\mu_{i}$ only depends on the marginal distributions of each $g_{k}$ separately, the program in Proposition \ref{prop-sensitivity} is equivalent to one that only has $K = 43$ variables.
}
We take the sensitivity set to be $\PM(\delta \vert \bar{q})$ where $\bar{q}$ puts marginal probability $8/30$ on Republican-led states and probability $11/13$ on the Democrat-led states.
These are the empirical ex-post probabilities of states expanding by party, which is what \cite{borusyakhull2026e} use to specify their shock distribution in their Monte Carlo simulations.
We abuse notation slightly by not applying the $\delta$ expansion in $\PM(\delta \vert \bar{q})$ to the Democrat-led states.\footnote{
  This would be like having $\delta_{k}$ depend on $k$ in the definition of $\PM$, with $\delta_{k}$ fixed at one for states $k$ that are Democrat-led.
}
This is intended to keep the exercise simple by considering sensitivity to the Republican-led states only.

\fig[\textwidth][p]%
{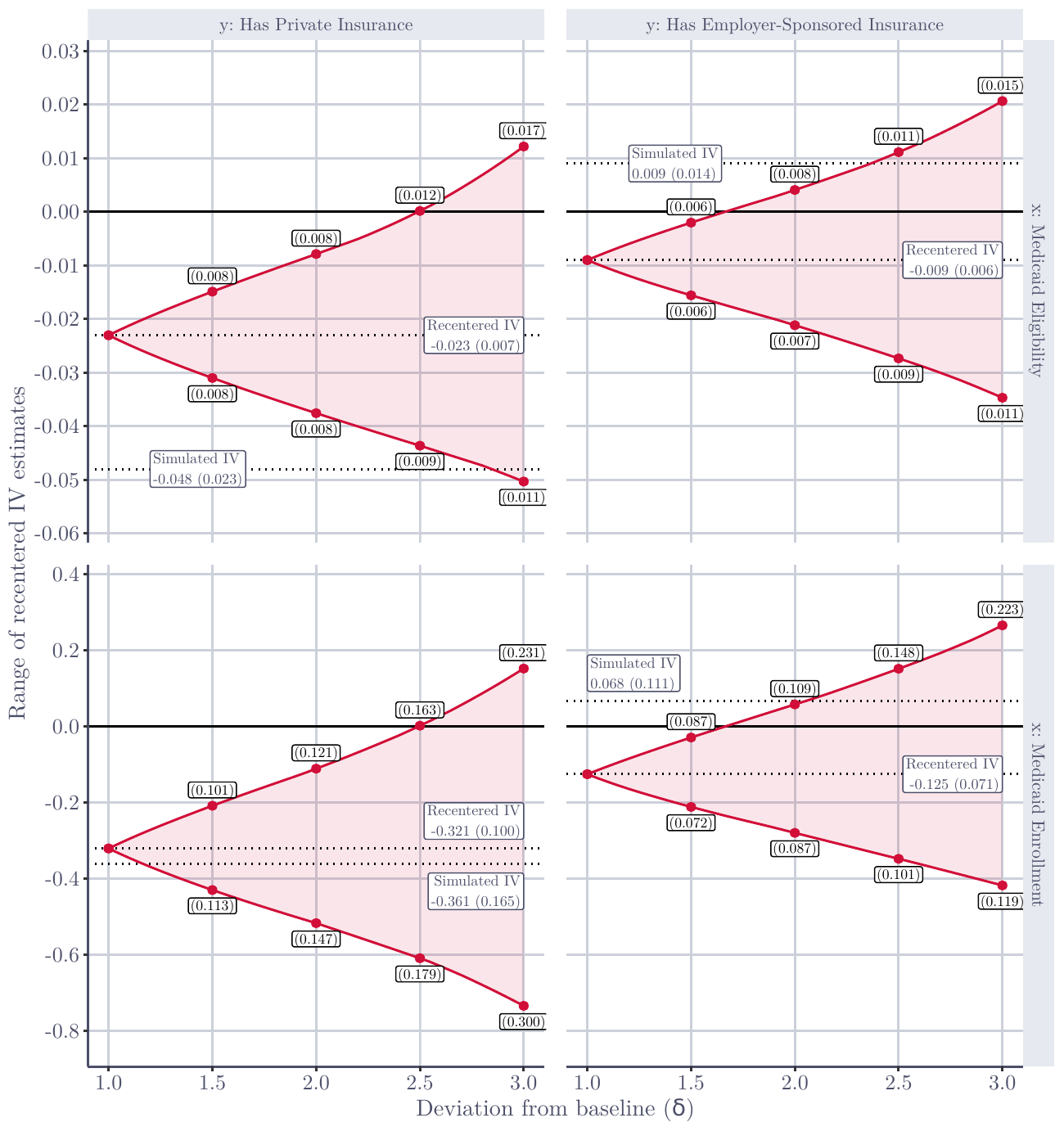}%
{fig:medicaid}%
{Sensitivity to heterogeneous Republican-led expansion probabilities}%
{%
  Bounds from solving \eqref{eq-sensitivity-lp} with $\mathcal{P} = \PM(\delta \vert \bar{q})$ for different values of $\delta$.
  Different outcomes $y_{i}$ are in the columns and different endogenous variables $x_{i}$ are in the rows.
  Following \cite{borusyakhull2026e}, we set the baseline $\bar{q}$ to be 8/30 for Republican-led states and 11/13 for Democrat-led states.
  However, we only consider sensitivity to allowing deviations from $\bar{q}$ for the Republican-led states.
  At $\delta = 1$, all Republican-led states have an equal ex-ante probability of expansion, which reproduces the \cite{borusyakhull2026e} recentered IV estimate.
  Standard errors clustered by state are shown in parentheses, following the same approach to inference as in \cite{borusyakhull2026e}.
  For the bounds, these show the standard errors at the optimizer for the program.
}

Figure \ref{fig:medicaid} shows the results for different values of $\delta$, together with the two instrumented difference-in-differences estimates reported by \cite{borusyakhull2026e}.
\cite{borusyakhull2026e} point out that the precision gains in their recentered IV estimate lead to standard errors for the impact of Medicaid eligibility on private insurance take-up that are 70\% smaller than for the simulated IV estimate.
The first row of Figure \ref{fig:medicaid} shows that this conclusion comes at the risk of bias from incorrectly specifying the expansion probabilities.
If Republican-led states are allowed to have ex-ante probabilities between $.11$ and $.67$ ($\delta = 2.5$) rather than a uniform $.27$ ($\delta = 1$), then a wide range of conclusions are available: eligibility could have a negative effect on private insurance take-up similar to that found by the simulated instrument or it could have a null effect.
The impacts on employer-sponsored health insurance are even more stark and show that the sign-flip found by \cite{borusyakhull2026e} is highly fragile to their assumed expansion probabilities.
The second row of Figure \ref{fig:medicaid} changes the endogenous variable from Medicaid eligibility to Medicaid enrollment, as in \cite{borusyakhull2026e} Table 2, Panel B.
Even greater sensitivity is found here; in particular the negative effect on employer-sponsored insurance can be statistically insignificant if ex-ante Republican-led expansion probabilities can vary between $.18$ and $.40$ ($\delta = 1.5$) and positive if these probabilities are allowed to vary between $.13$ and $.53$ ($\delta = 2.0$).

\section{Conclusion}
\label{sec-conclusion}

We developed a computationally tractable method for systematically assessing the sensitivity of estimators based on formula instruments to the assumed distribution of counterfactual shocks.
The estimator can be implemented in our companion package \texttt{formulaiv}.
We applied our estimator to both of the empirical applications in \cite{borusyakhull2023e} and \cite{borusyakhull2026e} and found both to exhibit substantial sensitivity to the assumptions on the distribution of counterfactual shocks.

Our analysis suggests that researchers using formula instruments should be cautious about the specification of counterfactual shocks.
When these shocks represent events such as the opening of a high-speed rail line or a state policy change, it seems like a challenging exercise to divine the ``correct'' shock distribution.
Other examples suggested in \cite{borusyakhull2023e}, such as the probability of earthquakes, likely face similar challenges, which can be assessed quantitatively using our methods.
These uses of formula instruments have begun to be adopted by empirical researchers: see, for example, \cite{dellolken2020res}, \cite{bosshartweigand2025}, \cite{buhlerdickens2025}, \cite{moroninicolettisalvanestominey2025}, and \cite{doellingsenlim2025wp}.

However, there are other uses of formula instruments that rely on institutional knowledge of how the instrument was assigned.
Examples include
\cite{chaureynayyarsharmaverhoogen2025}, \cite{hollenbeckhristakevauetake2025wp}, \cite{cailinszeidl2026wp}, \cite{baguesmakanyvattuonezinovyeva2026wp}, \cite{jensenkumarpoensgen2026wp}, and \cite{gao2026wp}.
Sensitivity to these formulas is likely a smaller concern, because the distribution of counterfactual shocks is determined by the randomization protocol.
For these applications, our method can be used to provide a robustness check to deviations from the stated protocol.

\inputbibliography{formulaiv}

\startappendix
\section{Proof of Proposition \ref{prop-sensitivity}}
\label{sec-proof}

We first establish the case where $D(p) \geq 0$ for all $p \in \mathcal{P}$.
The case where $D(p) \leq 0$ for all $p \in \mathcal{P}$ follows symmetrically after the noted changes.

The linear program \eqref{eq-sensitivity-lp} is the \cite{charnescooper1962nrlq} transform of the linear-fractional program \eqref{eq-fractional-lp}.
The two programs yield the same optimal values when $D(p) > 0$ for all $p \in \mathcal{P}$; see \citet[][pg. 151]{boydvandenberghe2004} for a textbook treatment.
If $D(p) = 0$ for some $p \in \mathcal{P}$, then the optimal value may be unbounded.
After the transformation, the simplex membership $p \in \Delta^{S}$ becomes $\phi \geq 0$ and $\sum_{s=1}^{S}\phi_{s} = \tau$, while the remaining constraints $Ap \leq c$ that define $\mathcal{P}$ become $A\phi \leq c\tau$.

Now suppose that $b \in [\betalb(\mathcal{P}), \betaub(\mathcal{P})]$ is a real number.
The objective function of \eqref{eq-sensitivity-lp} is continuous in $(\tau, \phi)$, and the constraint set of \eqref{eq-sensitivity-lp} is convex, hence connected.
So the image of the objective function over the constraint set is an interval with infimum $\betalb(\mathcal{P})$ and supremum $\betaub(\mathcal{P})$ \citep[e.g.][Theorem 4.22]{rudin1976}.
Because a feasible linear program attains any finite optimal value, this interval contains its finite endpoints and therefore contains every real $b \in [\betalb(\mathcal{P}), \betaub(\mathcal{P})]$.
It follows that there exists a $\phi(b), \tau(b)$ pair that is feasible in \eqref{eq-sensitivity-lp} that produces objective value $b$.
Suppose momentarily that $\tau(b) > 0$.
Let $p(b) = \phi(b)/\tau(b)$.
Then $p(b) \in \mathcal{P}$ and
\begin{align*}
  \betariv(p(b))
  &=
  \frac{
    \sum_{i=1}^{N}
    y_{i}z_{i}
    -
    \sum_{i=1}^{N}
    \sum_{s=1}^{S}
    y_{i}f_{is}p_{s}(b)
  }{
    \sum_{i=1}^{N}
    x_{i}z_{i}
    -
    \sum_{i=1}^{N}
    \sum_{s=1}^{S}
    x_{i}f_{is}p_{s}(b)
  } \\
  &=
  \frac{\tau(b)}{\tau(b)}
  \left(
    \frac{
      \sum_{i=1}^{N}
      y_{i}z_{i}\tau(b)
      -
      \sum_{i=1}^{N}
      \sum_{s=1}^{S}
      y_{i}f_{is}\phi_{s}(b)
    }{
      \sum_{i=1}^{N}
      x_{i}z_{i}\tau(b)
      -
      \sum_{i=1}^{N}
      \sum_{s=1}^{S}
      x_{i}f_{is}\phi_{s}(b)
    }
  \right)
  =
  b.
\end{align*}
This shows that there exists a $p(b) \in \mathcal{P}$ that produces $\betariv(p(b)) = b$.

We conclude the proof by showing that a feasible pair $\tau(b), \phi(b)$ cannot have $\tau(b) = 0$.
If $\tau(b) = 0$, then the constraints $\phi_{s}(b) \geq 0$ for all $s$ and $\sum_{s=1}^{S}\phi_{s}(b) = \tau(b) = 0$ force $\phi_{s}(b) = 0$ for all $s$.
But then the normalization constraint
\begin{align*}
  \sum_{i=1}^{N}x_{i}z_{i}\tau(b) - \sum_{i=1}^{N}\sum_{s=1}^{S}x_{i}f_{is}\phi_{s}(b) = 1
\end{align*}
reduces to $0 = 1$, contradicting the feasibility of $\tau(b), \phi(b)$.

Finally, suppose that $D(p)$ takes both positive and negative values on $\mathcal{P}$ and that $\betariv(p)$ is not constant on $\mathcal{P}_{D \neq 0}$.
Because $\mathcal{P}$ is connected and $D$ is continuous, there exists some $p_{0}$ such that $D(p_{0}) = 0$ while the numerator of $\betariv(p)$ is non-zero; suppose it is positive for concreteness.
Then taking a feasible sequence of $p$ that approaches $p_{0}$ from within the set $\{p \in \mathcal{P} : D(p) > 0\}$ produces arbitrarily large values of $\betariv(p)$, while taking a feasible sequence from within the set $\{p \in \mathcal{P} : D(p) < 0\}$ produces arbitrarily small values of $\betariv(p)$.
We conclude that $\betalb(\mathcal{P}) = -\infty$ and $\betaub(\mathcal{P}) = +\infty$.

\section{Extension to general assignment processes}
\label{sec-general-assignment}

In Section \ref{sec-sensitivity}, we assumed that $G(\cdot \vert w)$ is independent of $w$ with discrete support.
In this appendix, we relax this assumption by assuming instead that $G(\cdot \vert w)$ has a density $\gamma(\cdot \vert w)$ with respect to some known dominating measure $\lambda$.
Then
\begin{align}
  \label{eq-general-mu-expression}
  \mu_{i}
  \equiv
  \Exp[f_{i}(g, w) \vert w]
  =
  \int f_{i}(g,w) \gamma(g \vert w)\, d\lambda(g).
\end{align}
Suppose that $\gamma$ can be written using a finite basis expansion as
\begin{align}
  \label{eq-basis-expansion}
  \gamma(g \vert w)
  =
  \sum_{s=1}^{S}
  p_{s}\gamma_{s}(g, w),
\end{align}
where $\gamma_{s}$ are known basis functions.
This nests the case considered in the main text by taking $\lambda$ to be counting measure on the finite set $\{g_{1},\ldots,g_{S}\}$ and $\gamma_{s}(g \vert w) = \IndicSmall{g = g_{s}}$ for $s = 1,\ldots,S$.
Substituting \eqref{eq-basis-expansion} into \eqref{eq-general-mu-expression} produces
\begin{align}
  \mu_{i}
  =
  \sum_{s=1}^{S}p_{s} \underbrace{\int f_{i}(g,w)\gamma_{s}(g,w)\, d\lambda(g)}_{f_{is}}
  \equiv
  \sum_{s=1}^{S}p_{s}f_{is},
\end{align}
with $f_{is}$ redefined (generalized) from the main text.
This has the same form as equation \eqref{eq-mu-as-discrete-sum}, except that instead of being known directly, the quantities $f_{is}$ need to be computed, for example by drawing from $\lambda(g)$.
Having done that, Proposition \ref{prop-sensitivity} proceeds unchanged as long as $p$ is constrained to a polyhedral sensitivity set $\mathcal{P}$.

\section{Predicting HSR line openings}
\label{sec-sa-predicting-marginal-shock-probs}

In this section, we describe how we use machine learning algorithms to predict the opening probability of HSR lines for the application in Section \ref{sec-china}.

The variable being predicted is a binary indicator for whether the HSR line was open by 2016.
The predictors are variables predetermined as of 2007: the 2007 opening status, anticipated railway speed, line length, line type, number of links, and plan type.
A few plan type categories appear for only one or two lines, which we pool into a separate ``other'' category.
\cite{borusyakhull2023e} set some line opening probabilities to one across all of their scenarios; we continue to do this in our prediction exercise, while focusing our attention on the other lines.

\fig%
{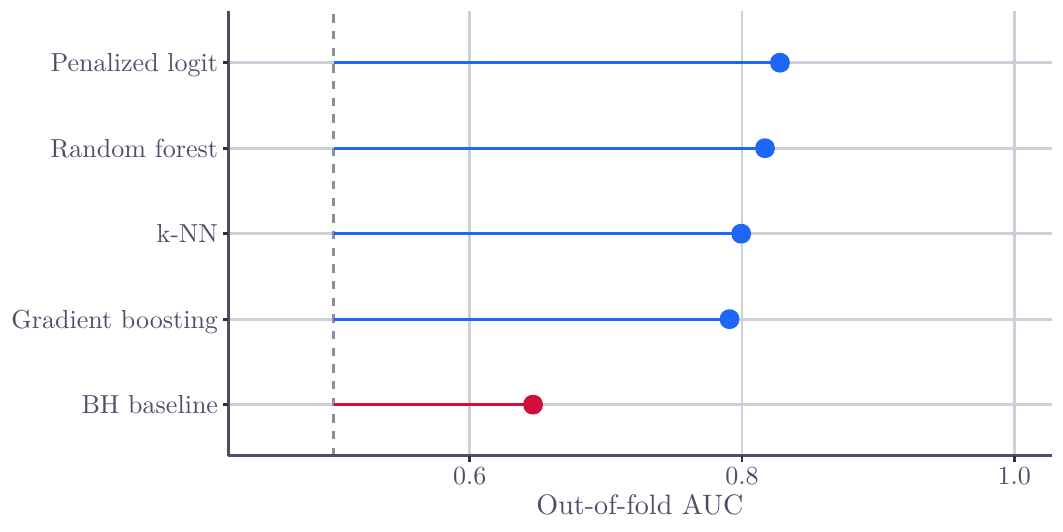}%
{fig:model_eval}%
{Out-of-fold forecast performance}%
{%
  See the text of Appendix \ref{sec-sa-predicting-marginal-shock-probs} for details.
}

We train four different learners: a random forest, penalized logistic regression, gradient-boosted trees, and $k$-nearest neighbors.
For each one, we use nested cross-validation and evaluate performance using the area under the ROC (true positive rate/false positive rate) curve, often abbreviated as the AUC.
In the outer loop, we split the data into five folds.
In the inner loop, we use ten-fold cross-validation within the four training folds (repeated five times) to select tuning parameters by cross-validated AUC.
We then use the optimal tuning parameters to construct a prediction for lines in the left-out fifth fold.
Repeating this process for each of the five folds produces out-of-fold predictions for each line.

Figure \ref{fig:model_eval} reports the out-of-fold AUC for each learner together with the AUC for the naive data-agnostic predictions implied by the \cite{borusyakhull2023e} baseline $\qbh$.
Unsurprisingly, the learners that use data perform substantially better with out-of-fold AUC between roughly $.79$ and $.83$, compared to $.65$ for the $\qbh$ baseline.
As we saw in Figure \ref{fig:histogram_ml} in the main text, the four learners produce substantially different marginal distributions $\bar{q}_{m}$ even while performing comparably on the AUC measure.

\clearpage
\section{Additional figures}
\label{sec-sa-additional-figures}

\fig[\textwidth][h!]%
{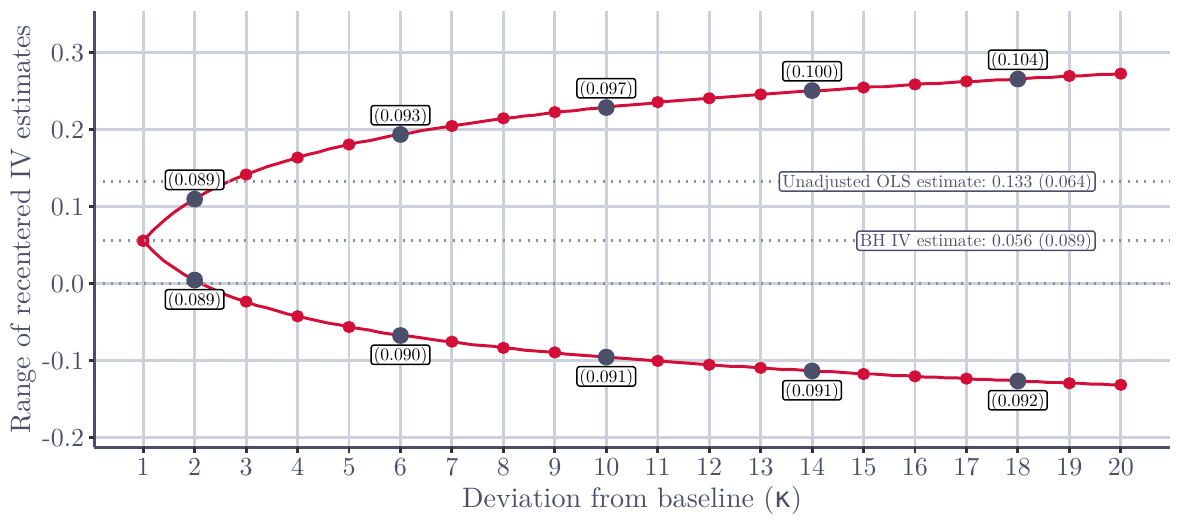}%
{fig:est_joint_with_controls}%
{Adding geographic controls to Figure \ref{fig:est_joint_no_controls}}%
{%
  The figure is the same as Figure \ref{fig:est_joint_no_controls} but with controls for distance to Beijing, latitude, and longitude, as in Panel B of Table I in \cite{borusyakhull2023e}.
  See notes for Figure \ref{fig:est_joint_no_controls}.
}

\end{document}